# Reliable determination of the Cu/n-Si Schottky barrier height by using in-device hot-electron spectroscopy

Subir Parui[1,*], Ainhoa Atxabal[1], Mário Ribeiro[1], Amilcar Bedoya-Pinto[1],

Xiangnan Sun[1], Roger Llopis[1], Fèlix Casanova[1,2], Luis E. Hueso[1,2,*]

[1]CIC nanoGUNE, 20018 Donostia-San Sebastian, Basque Country, Spain

[2]IKERBASQUE, Basque Foundation for Science, 48011 Bilbao, Basque Country, Spain

Email: s.parui@nanogune.eu; l.hueso@nanogune.eu

## Abstract

We show the operation of a $Cu/Al_2O_3/Cu$/n-Si hot-electron transistor for the straightforward determination of a metal/semiconductor energy barrier height even at temperatures below carrier-freeze out in the semiconductor. The hot-electron spectroscopy measurements return a fairly temperature independent value for the Cu/n-Si barrier of $0.66 \pm 0.04$ eV at temperatures below 180 K, in substantial accordance with mainstream methods based on complex fittings of either current-voltage ($I$-$V$) and capacitance-voltage ($C$-$V$) measurements. The Cu/n-Si hot-electron transistors exhibit an OFF current of $\sim 2 \times 10^{-13}$ A, an ON/OFF ratio of $\sim 10^5$ and an equivalent subtreshold swing of $\sim 96$ mV/dec at low temperatures, which are suitable values for potential high frequency devices.

The Schottky energy barrier naturally appearing at a metal/semiconductor (MS) interface is a critical parameter for the performance of many modern electronic devices, from mainstream metal-oxide-semiconductor field effect transistors (MOSFETs) to novel organic-based light-emitting diodes or photovoltaics [1-5]. Hot electron in-device spectroscopy is a powerful technique for determining such energy barrier between a metal and a semiconductor [6-9]. This technique is based in the hot-electron transistor (HET), a classical metal-A/insulator/metal-B/semiconductor (MIMS) device in which a hot electron current coming from a metal-A emitter crosses ballistically a metal-B base before entering into the semiconductor conduction band. The HET is particularly interesting since hot electrons can be collected in the semiconductor without biasing the MS interface, and hence they generate a collector current that can be used to probe the Schottky barrier without the effects of an external electric field at the interface. Moreover, the energy barrier height can be obtained directly from the experimental data with a simple theoretical model [10]. However, and in spite of its power and simplicity, hot-electron spectroscopy has been sparsely used for the study of metal/inorganic-semiconductor [6, 7, 10-14] and, more recently, metal/organic-semiconductor interfaces [8-9].

In this manuscript, we explore the prototypical Cu/n-Si interface and extract its energy barrier by hot-electron spectroscopy. Additionally, our HETs exhibit appealing electronic characteristics such as an OFF current of $\sim 2 \times 10^{-13}$ A, an ON/OFF ratio of $\sim 10^5$, and an equivalent subthreshold swing of ~96 mV/dec. The Cu/n-Si combination is particularly interesting since copper grows highly textured on Si [14, 15], has a high electrical conductivity and strong electromigration resistance, properties which make it a model system for investigation compared to other polycrystalline interfaces.

Previously, hot electron spectroscopy have been used for exploring the attenuation length of hot electrons in Cu [16, 17], identifying the band structure of the underlying n-Si substrate [14] and the crystallographic orientation of n-Si [16], detecting *chemicurrent* [18], as well as for spintronic applications [14, 19-20]. However, in our HET device, we capture useful information such as the reduction of thermal-leakage current with decreasing temperature [21] in addition to the reduction of quasi-elastic phonon scattering of hot electrons in the metal base and in the semiconductor collector [22]. Most importantly, we provide reliable determination of the Schottky barrier of Cu/n-Si interface at low temperatures even below carrier freeze-out temperature of Si, a regime that can be hardly accessed by other conventional electrical characterization methods. For putting this technique into the right perspective, we compare the results we obtained by hot-electron spectroscopy with those arising from standard current – voltage (*I-V*) and capacitance – voltage *(C-V)* measurements taken on the same device. Our work underlines in-device hot-electron spectroscopy as a method of choice to map out the Schottky barrier physics below carrier freeze-out temperatures in a wide range of semiconducting materials.

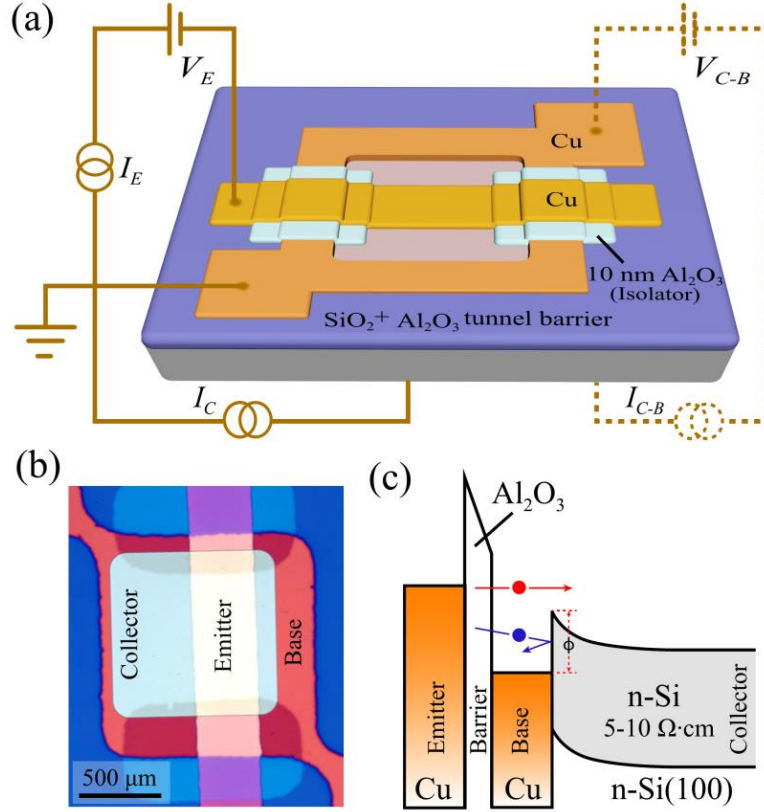

Figure 1: (a) Schematic diagram of the hot-electron transistor. The dotted line represents the *I-V* measurement setup to measure the base/collector diode. (b) Optical microscopy image of the real device. (c) Schematic energy diagram of the hot electron transistor, representing the negatively biased Cu emitter, the $Al_2O_3$ tunnel barrier, and the grounded Cu base in direct contact with n-Si(100) collector. The base/collector junction is unbiased.

Figure 1(a) shows the schematics of the $Cu/Al_2O_3/Cu$/n-Si HET device. The collector used is a n-type Si(100) substrate, having resistivity of 5-10 Ω·cm (donor concentration, $N_D \approx 10^{15}$ $cm^{-3}$ at room temperature), with 300-nm-thick thermally grown $SiO_2$ on top. A 1×1mm$^2$ window was opened on the oxide layer by means of photolithography and buffered hydrofluoric acid etching. The etched area was then hydrogen terminated using 1% hydrofluoric acid, onto which a 20-nm-thick Cu base layer was deposited by e-beam evaporation in ultrahigh vacuum (UHV) through a shadow mask. For the hot-electron injection by quantum tunneling, a 3-nm-thick Al layer was then evaporated all over the device without breaking the vacuum and *in-situ* plasma oxidized, followed by another 20-nm-thick Cu emitter layer deposition through a different shadow mask. Prior to the evaporation of the second Cu layer, and to avoid voltage breakdown up to an emitter voltage $|V_E| \approx 1.4$ V of the thin tunnel barrier at the edges of 300-nm-thick $SiO_2$, an additional 10-nm-thick $Al_2O_3$ layer was evaporated in the area and surroundings of the Cu layers with the use of another shadow mask [see Fig. 1(a)]. Figure 1(b) pictures the top view of an actual device, with the emitter, base and collector indicated. The active area for hot-electron injection and collection is considered to be confined to the overlapping area between the emitter and base electrodes ($400 \times 800$ $\mu m^2$). Once fabricated, the device was then transferred into a variable-temperature probe station (Lakeshore) for electrical measurements with a Keithley-4200 semiconductor analyzer. Figure 1(c) represents the energy diagram of the HET device. The arrows represent the hot-electron transport above the Cu/n-Si barrier (red) and the electron reflection at the barrier interface below the barrier (blue), respectively. At a sufficiently high negative emitter bias ($-V_E$), the electrons tunnel through the barrier and cross the metal base ballistically with enough energy to surpass the base/collector Schottky barrier and get collected into the semiconductor, providing the collector current. The onset of the output characteristic of the device represents accurately in a first approximation the height of the Schottky energy barrier. Those electrons that do not have enough energy to surpass the Cu/n-Si energy barrier are scattered/reflected at the base/collector interface and contribute to the base current directed to ground.

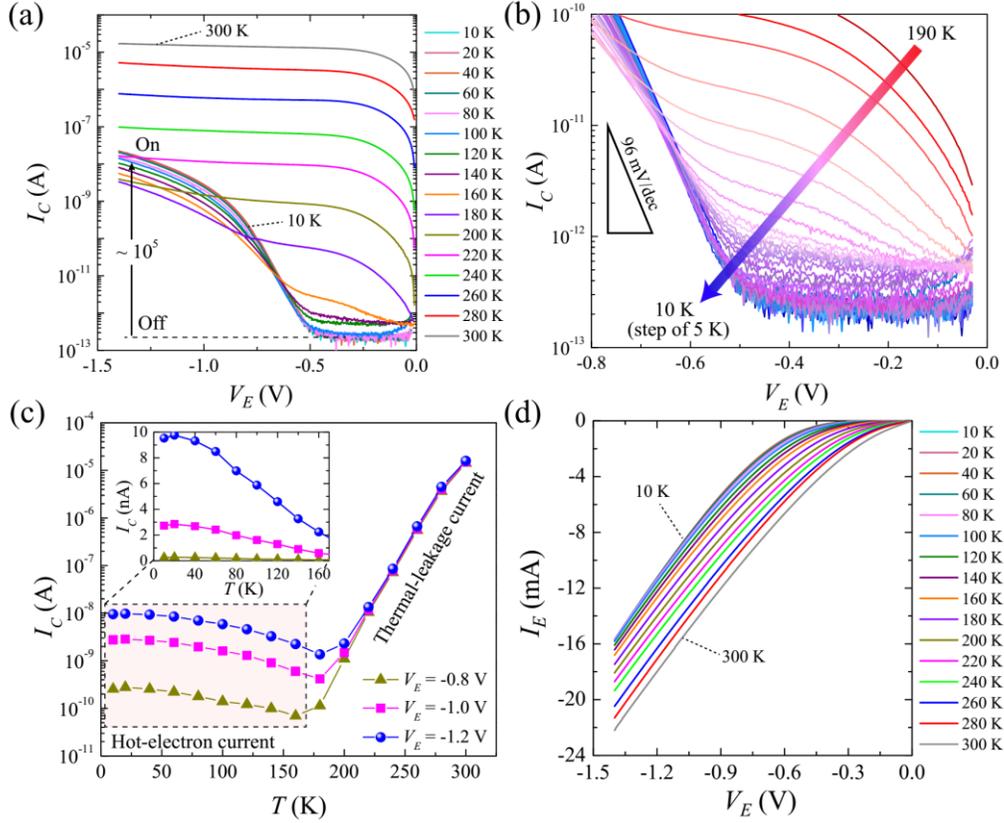

Figure 2: (a) Plots of the collector current ($I_C$) versus emitter bias ($V_E$) at different temperatures for grounded base and at zero collector bias. (b) $I_C$-$V_E$ plots at different temperatures with 5 K step. (c) $I_C$-$T$ plots at different $V_E$, which describe the reduction of thermal leakage current and collection of hot-electron current. Inset shows the collector current only due to hot-electrons. (d) Plots of the emitter current ($I_E$) versus emitter bias ($V_E$) at different temperatures.

In order to investigate the collector current originated from the ballistic transport of hot electrons through the Cu-base, we measure the temperature dependence of the collector current ($I_C$) versus emitter voltage ($V_E$) from 300 K down to 10 K (see figure 2). As noted above, the onset of the collector current represents an adequate measurement for the energy barrier at the Cu/n-Si interface. At 300 K, a slowly varying collector current of ~ $10^{-5}$ A is observed without any well-defined onset above $V_E$ = -0.3 V, as it can be seen in Figure 2 (a). The magnitude of this collector current decreases as temperature does, confirming that its origin is the regular thermal leakage current at the Cu/n-Si Schottky interface [21]. For temperatures below 180 K, the hot-electron current (ON) arises for $V_E$ higher than -0.5 V, reaching values as high as ~ $10^{-8}$ A, while the OFF current (for $V_E$ lower than -0.5V) is as low as ~$2 \times 10^{-13}$ A. The OFF current measured is around two orders of magnitude lower than the reported in recent literature [20] and, accordingly, our HET devices exhibit a large ON/OFF ratio of ~$10^5$. The equivalent subthreshold swing (defined as the required $V_E$ for a ten-fold change in $I_C$) is ~96 mV/dec at ~80 K and below, a very suitable value for applications in high-speed and frequency integrated-circuits [24, 25]. Figure 2 (b) shows in detail (in 5 K steps) the evolution of the hot electron current for temperatures below 190 K, while Figure 2 (c) displays an exponential decrease with decreasing temperature for the thermal leakage down to 160 ± 20 K. However, at temperatures below 160 K, the hot-electron current increases linearly for $V_E$ voltages larger than the metal/semiconductor barrier height. Finally, for temperatures below ~20 K the hot electron current decreases due to the carrier freeze out in n-Si where the semiconductor becomes highly resistive (see Figure 2 (c), inset) [26]. Figure 2 (d) represents the tunnel current vs emitter bias at several temperatures. Observation of a large amount of emitter current and also the linearity at high emitter bias suggest the possibilities of unavoidable current through the pinholes in the barrier that might originated from the island growth of Al on top of Cu. However, the thin amorphous $Al_2O_3$ barrier in our devices is good enough for the injection of energetic electron from the Cu-emitter into the Cu-base across the MIM structure. The emitter current slowly decreases with decreasing temperatures, whereas the hot-electron current increases, possibly due to the reduction of quasi-elastic acoustic-phonon scattering in the base as well as in the collector at low temperatures [22]. As seen from this set of results, we describe the evolution of the collector current from 300 K to 10 K to interpret the origin of hot electron current and thermal-leakage current, while some previous experiments of hot electron transport across Cu/n-Si interface are reported at temperatures below 150 K [14, 16, 17, 19, 20, 27].

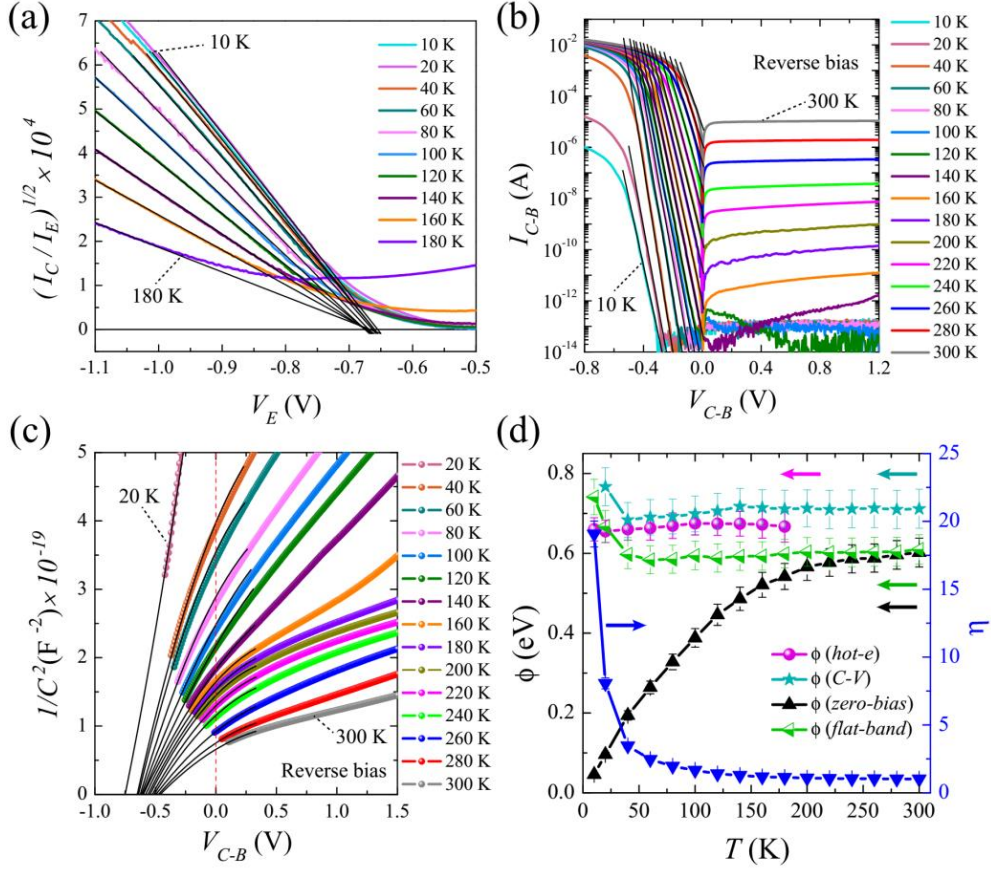

Figure 3: (a) Hot-electron current plotted as $(I_C/I_E)^{1/2}$ versus $V_E$ superimposed to the linear fit (solid line) at several temperatures. Linear fits are used to extract the barrier heights. (b) Temperature dependent base-collector I-V diode characteristics in semi-logarithmic scale. (c) Base-collector $C^{-2}$-V characteristics of the diode at a frequency of 1 MHz and in the temperature range from 300 K to 20 K. Solid lines represent the extrapolation of the curvature of the experimental data. (d) Temperature dependence of the barrier heights extracted by different experimental methods i.e., hot-electron, C-V, and I-V (zero-bias and flat-band); measurements are plotted in the left vertical axis. Right vertical axis represents the change in ideality factor of the diode.

For the reliable determination of the Schottky barrier at the Cu/n-Si interface, we compare the values extracted from hot-electron spectroscopy at several temperatures with the ones inferred from *I-V* and *C-V* methods of the same base/collector junction [see Figure 3]. In hot-electron spectroscopy, the barrier height is extracted as the intercept with the voltage axis for zero collector current. Due to the parabolic conduction band minimum in n-Si, the theoretical dependence (Bell and Kaiser model) of $I_C$ with $V_E$ [10] is $I_C \propto I_E(V_E - \phi)^2$, where ϕ is the barrier height of the metal base/semiconductor collector interface. Figure 3 (a) shows the hot-electron spectroscopy data, where $(I_C/I_E)^{1/2}$ is plotted with respect to $V_E$ and fitted linearly to the experimental curve. According to Schottky-Mott relationship, the value of ϕ is typically the difference between the base-metal work function and the conduction band edge of n-Si(100) collector in the ideal case. However, in general, the energy-level alignment is influenced by additional effects such as Fermi-level pinning, so that the rigid-band approximation is no longer suitable for capturing the actual barrier height [1, 2, 23]. Figure 3 (d) displays experimentally extracted barrier heights by hot-electron spectroscopy and we can observe that its value below 180 K is constant around ~ 0.66 ± 0.04 eV.

We can now compare the hot-electron spectroscopy data with other mainstream techniques for obtaining the energy barrier height. In the first place, we show two-probe *I-V* measurements of the base/collector (metal/semiconductor) junction [see Figure 1 (a) for the setup and Figure 3 (b) for the results]. The data shows clearly the rectifying characteristics of a MS diode. The rectification factor reaches several orders of magnitude and it increases with decreasing temperature due to the sharp decrease of the reverse-bias saturation current ($I_{sat}$). However, below 40 K the forward saturation current starts to decrease due to the fact that the Si-collector becomes resistive because of carrier freeze-out in Si. The amplitude and temperature dependence of the reverse saturation current varies in a similar way as the hot-electron leakage current, which confirms that the reverse saturation current of the base/collector diode contributes to the leakage current. Using the linear part of the forward bias regime, we extract the barrier height after a fitting with the thermionic emission (TE) equation [1, 2] according to which $I_{C-B} = I_{sat}\left[exp\left(\frac{qV_{C-B}}{\eta k_B T}\right)\right]$ for $V_{C-B} > 3k_B T/q$, where $k_B$ is the Boltzmann constant, $T$ is the temperature and $\eta$ is the ideality factor. Furthermore, $I_{sat} = AA^*T^2 exp\left(-\frac{q\phi_{b0}}{k_B T}\right)$, where $A$ is the base/collector junction area and $A^*$ is the effective Richardson constant (~110 A cm$^{-2}$ K$^{-2}$ for n-Si) [1]. In principle, $\phi_{b0}$ is meant to be the same Cu/n-Si Schottky barrier as determined by hot electron spectroscopy; however, due to the different measurement methods we denote this as zero-bias barrier height [1] and plot the extracted barrier height in Figure 3 (d) along with the ideality factor $\eta$, a quantity which should be close to 1 for a well-defined Schottky interface. The zero-bias barrier height obtained from the $I_{C-B}$-$V_{C-B}$ measurement decreases monotonically from 0.60 ± 0.04 eV at 300 K to 0.046 ± 0.003 eV at 10 K, whereas the ideality factor sharply increases from 1.02 ± 0.04 at 300 K to 19.06 ± 0.95 at 10 K. These phenomena are commonly observed in Schottky diodes [28-30], and can be explained by considering an inhomogeneous distribution of barrier heights at the MS interface [31, 32] only in a temperature regime without the carrier freeze-out in the semiconductor. Here, such strong disagreement between these two measurement methods suggests that the zero-bias barrier height decreases artificially at low temperatures below 60 K, where the diode becomes extremely non-ideal and the thermionic emission model fails to explain the experimental data. Furthermore, taking into account the zero-bias barrier height and the ideality factor, the $\phi_{bf}$ (flat-band barrier height) [33] can be expressed as $\phi_{bf} = \eta\phi_{b0} + (\eta - 1)(k_B T/q)[\ln(N_C/N_D)]$. The extracted $\phi_{bf}$ is plotted in Figure 3(d) representing the correction of the non-ideal $\phi_{b0}$ to achieve a more realistic Schottky barrier height. In addition, it is also possible to obtain a Schottky barrier height from the Richardson plots using the reverse bias diode characteristics [2, 4, 34]. We find the Cu/n-Si Schottky barrier height to be $0.52 \pm 0.02$ eV, fitting the $I_{sat}$ current only in the high temperature regime (300 K to 200 K). However, this method of extracting the Schottky barrier height underestimates the influence of the diode ideality factor.

In the second place, we show the capacitance (C)-voltage (V) measurements and how we can extract in this case again the metal/semiconductor energy barrier, which we will denote as $\phi$ (C-V) [see Figure 3 (c)]. The applied voltages in this particular case range from the reverse-bias condition to a small regime of forward-bias until the diode starts conduction i.e., maximum up to -0.36 V. In the simplest case of a linear relation between $C^{-2}$ and $V$ we can write [2], $1/C^2 = (2/qN_D\varepsilon_s A^2) \cdot [V_I + V_{C-B}]$, where $V_I = (\phi - k_B T/q - \xi)$ represents the linear intercept on the voltage axis and $\varepsilon_s$ is the permittivity of Si. In case of linear dependence, the barrier height can be extracted as $\phi = V_I + k_B T/q + \xi$, where $\xi = (k_B T/q)[\ln(N_C/N_D)]$ is the energy difference between the Fermi level and the conduction band edge in Si. The effective density of states in the conduction band edge is given by $N_C \approx 2.8 \times 10^{19}$ cm$^{-3}$[1]. However, we notice a strong nonlinear dependence of the $C^{-2}$-$V_{C-B}$ characteristics. In case of nonlinearity, the $C^{-2}$ can be expressed as a quadratic function i.e., $1/C^2 = M_1[V_I + V_{C-B}] + M_2[V_I + V_{C-B}]^2$, where $M_1$ and $M_2$ are the fitting parameters [35]. Such nonlinear dependence could be due to a contribution of the interface charge capacitance that goes in adverted by other experimental techniques. By using a quadratic dependence around zero-bias voltage, we extract the intercept on the voltage axis, which is then used to determine the barrier as shown in Figure 3 (d). It is worth noting that, in our measurement, the $C^{-2}$-$V_{C-B}$ characteristics strongly deviate at 20 K, which could be possibly due to carrier freezing in n-Si at ~20 K and below.

Figure 3 (d) shows the different values of the barrier height extracted by hot electron spectroscopy, $I$-$V$ and $C$-$V$ measurements, i.e., $\phi$(hot-electron), $\phi$ (zero-bias), $\phi$ (flat-band) and $\phi$ $(C-V)$, respectively. They all show a similar weak temperature dependence, however $\phi$ (flat-band) and $\phi$ $(C-V)$ diverge at very low temperatures, while $\phi$(hot-electron) remains stable for all the temperatures obtained. Consequently, the hot-electron spectroscopy provides a reliable and straightforward method for the barrier height extraction in the absence of an applied electric field at the Cu/n-Si interface and at temperatures below the carrier freeze-out. Being an in-device method, as opposed to photoemission spectroscopic techniques, it provides a realistic approach to the metal/semiconductor barrier height determination for its application in electronic device design. Finally, we summarize the measured Schottky barrier heights by different methods in Table I, where we find a relatively good agreement with the existing literature [16, 17, 36] only in some certain temperatures.

TABLE I. Measured Schottky barrier heights by different methods and its comparison with the literature.

| T | $\phi$ (hot-e) | $\phi$ (C-V) | $\phi_{bo}$ (I-V) | $\phi_{bf}$ | $\phi$ (hot-e) from Ref. 16 | $\phi$ (hot-e) from Ref. 17 | $\phi_{b0}$ (I-V) from Ref. 35 |
|---|---|---|---|---|---|---|---|
| 300 K | - | 0.71 eV | 0.60 eV | 0.61 eV | - | - | 0.59 eV |
| 200 K | - | 0.71 eV | 0.56 eV | 0.60 eV | - | - | 0.61 eV |
| 100 K | 0.67 eV | 0.70 eV | 0.39 eV | 0.59 eV | 0.62 eV | - | 0.63 eV |
| 80 K | 0.67 eV | 0.69 eV | 0.33 eV | 0.58 eV | - | 0.64 eV | - |
| 40 K | 0.66 eV | 0.68 eV | 0.19 eV | 0.60 eV | - | - | - |
| 10 K | 0.66 eV | - | 0.05 eV | 0.74 eV | - | - | - |

In conclusion, we have demonstrated the operation of a Cu/n-Si-based hot-electron transistor and determined the height of its base/collector (metal/semiconductor) Schottky energy barrier. In this experiment we are able to fully describe the collector current that originates from hot-electron transport and to extract the Schottky barrier height over a wide temperature range. For completeness, we have compared the extracted barrier height with the other measurements methods such as the *I-V* and the *C-V* techniques, concluding that hot-electron spectroscopy is the most reliable method to map out Schottky barrier heights at very low temperatures. In addition, our results highlight in-device hot-electron spectroscopy as a straightforward method to determine the metal (base)/semiconductor (collector) energy barrier in real device operation conditions. Our optimized hot-electron transistor, with a current ON/OFF ratio of up to five orders of magnitude, low OFF state current, and an equivalent subthreshold swing of ~96 mV/dec at low temperatures, could be suitable for high-frequency device applications in cryogenic environments.

We acknowledge P. Stoliar for his help in the shadow mask design and for discussion regarding the capacitance measurements. We also acknowledge financial support from the European Union's 7th Framework Programme under the European Research Council (Grant 257654-SPINTROS), under People Programme (Marie Curie Actions) REA grant agreement 607904-13, and under the NMP project (NMP3-SL-2011-263104- HINTS) as well as by the Spanish Ministry of Economy under Project No. MAT2012-37638.